\documentclass[compress,two column,3p]{elsarticle}

\usepackage{lineno}
\usepackage{amssymb,amsmath} %for align

\usepackage{graphicx}% Include figure files
\usepackage{dcolumn}% Align table columns on decimal point
\usepackage{bm}     % bold math

\usepackage{xspace}             % espaces automatiques en fin de macro

\usepackage{booktabs}

\usepackage{hyperref}

\usepackage[justification=justified]{caption}
\usepackage{url}
\journal{IJMS: Accepted for publication}

\begin{document}

\begin{frontmatter}

%\preprint{}

%\title{ElectroSpray Ion Source for mass calibration of the Multi-Reflection Time-of-Flight Mass Spectrograph dedicated to Mass Measurement of short-lived nuclides}
 % \title{Production of mass calibrants for superheavy elements using  electrospray ionization with an rf-carpet }
  %  \title{An rf-carpet electrospray ion source for the calibration of an MRTOF Mass Spectrograph to be used in superheavy elements mass measurements}

 \title{An rf-carpet electrospray ion source to provide isobaric mass calibrants for trans-uranium elements}

\author[riken]{S.~Naimi}
\author[tsukuba,riken]{S.~Nakamura}
\author[tsukuba,riken]{Y.~Ito}
\author[tsukuba,riken]{H.~Mita}
\author[sophia]{K.~Okada}
\author[tsukuba]{A.~Ozawa}
\author[tsukuba,riken]{P.~Schury}
\author[riken]{T.~Sonoda}
\author[Aoyama]{A.~Takamine}
\author[riken]{M.~Wada}
\author[geissen]{H. Wollnik}

 \address[riken]{SLOWRI Team, Nishina Accelerator-based Research Center, RIKEN, 2-1 Hirosawa, Wako, Saitama 351-0198, Japan}
\address[tsukuba]{University of Tsukuba, 1-1-1 Tennodai, Tsukuba, Ibaraki 305-8577, Japan} 
\address[sophia]{Sophia University, 7-1 Kioi-cho, Chiyoda-ku, Tokyo 102-8554, Japan}
\address[Aoyama]{Department of Physics and Mathematics, Aoyama Gakuin University, 5-10-1 Fuchinobe, Chuo, Sagamihara, Kanagawa 252-5258, Japan}
\address[geissen]{New Mexico State University, Departement Chemistry and Biochemistry, Las Cruces, NM 88003,USA}

\date{\today}

\begin{abstract}
For trans-uranium elements, stable atomic isobars do not exist. 
In order to provide isobaric reference ions for the mass measurement of trans-uranium elements, an electrospray ion source (ESI) was combined with an rf-carpet to collect molecular ions efficiently. 
The rf-carpet allows for simplification of the pumping system to transport ions from the ESI to a precision mass analyzer. 
Molecular ions appropriate for isobaric references of trans-uranium elements were extracted from the rf-carpet and analyzed by a multi-reflection time-of-flight mass spectrograph (MRTOF-MS) with a resolving power of $\rm{R_m} \gtrsim100,000$.
\end{abstract}


\end{frontmatter}
%\pacs{Put Something!!!!!!!}

%\maketitle

\section{Introduction}

Many chemical elements heavier than uranium have been discovered over the last few decades. 
There has been tremendous  progress towards production of ever heavier elements with the goal of discovering the so-called superheavy element (SHE) `islands of stability' \cite{Heenen2002,Hofmann2000}. 
The stability of a nucleus essentially depends  on its binding energy, which defines how strongly the nucleons are bound together and  reflects the shell effects \cite{Lunney2003} that play a key role in stabilizing  trans-uranium elements against spontaneous fission \cite{A.Sobiczewski1966}.  
As yet, in the entire history of the SHE quest, only five masses of trans-uranium elements have been measured directly, by the SHIPTRAP Penning trap mass spectrometer at GSI \cite{Block2010,Dworschak2010,Ramirez2012}. 
All other known trans-uranium element's masses have been determined indirectly via $\alpha$-decay, which is not always reliable, especially for odd nuclei where, in most cases, the $\alpha$-decay includes excited states.  
Moreover, error propagation leads to a compounding of errors along the decay chain. 
Thus, for an unambiguous determination of the binding energy, a direct mass measurement is very important, which also determines the reaction Q-values, important for efficient synthesis of trans-uranium elements via fusion reactions \cite{Hofmann2000}. 

Recently, we have developed  a multi reflection time-of-flight mass spectrograph (MRTOF-MS) for mass measurement of short-lived nuclei  \cite{Wollnik1990,Ishida2004,Ishida2005,Schury2009}. 
This device can be also used as an isobar separator of short-lived nuclei as is the case for the MRTOF mass separator at the ISOLTRAP spectrometer at ISOLDE/CERN \cite{Wolf2012}.  
In the near future, it is planned to place our MRTOF-MS at the gas-filled recoil separator GARIS at the RIKEN linear accelerator facility RILAC, a world leading synthesizer of  trans-uranium elements \cite{Morita2004,Morita2004a,Morita2012}.  
To determine the mass of a given ion at least two known reference masses are required; the reference masses should, preferably, be isobars of the species of interest. 
However, trans-uranium elements do not have stable \textit{atomic} isobars that can be used as references. 
Therefore, we have built a new electrospray ion source \cite{Yamashita1984,Fenn1989,Fenn1993,Kebarle2009}, with improved vacuum coupling, to deliver isobaric \textit{molecular} ions as reference ions for trans-uranium elements.  
In this ion source, the conventional skimmer is replaced by a radiofrequency carpet (rf-carpet) \cite{Wada2003,Takamine2005} for a more efficient transport of molecular ions into the high vacuum region. 
The small exit hole negates the need for a sophisticated pumping system, while the simple geometry provides efficient extraction of ions. 
In addition, the use of the rf-carpet allows efficient transport of low-abundance ions.

\section{Experimental setup}

\begin{figure}[!htb]
\centering
\includegraphics[width=\columnwidth]{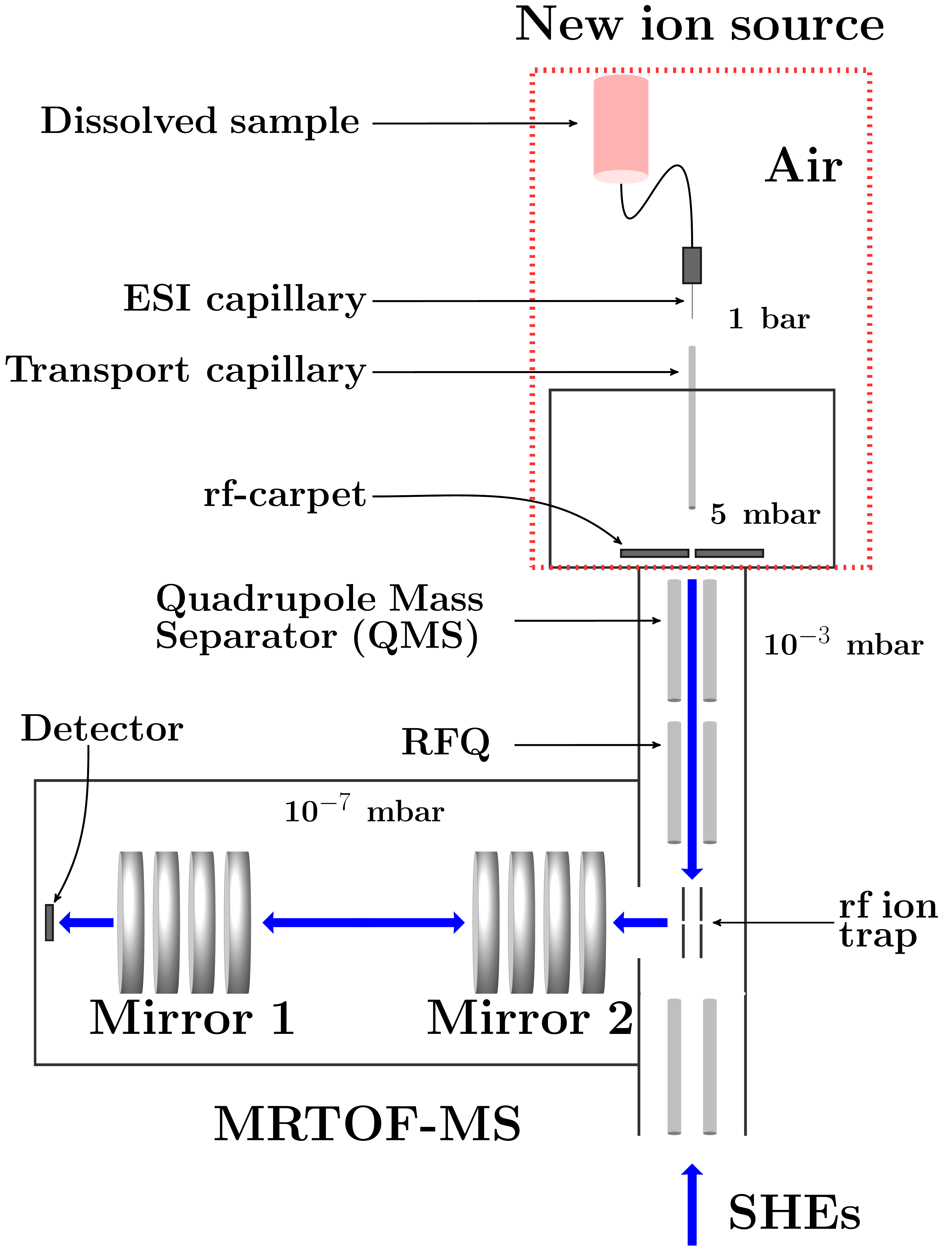}
\caption{(Color online) Schematic overview of the experimental setup. This setup will be placed at GARIS, where SHE beams could be collected from the side of the rf ion trap opposite to the ion source. }%
\label{fig:MTOFESI}%
\end{figure}

The experimental setup consists of the newly developed ion source and the MRTOF-MS combined with an accumulation-cooler rf ion trap (\autoref{fig:MTOFESI}). 
The ion source has an electrospray ionization (ESI) system for ion production and an rf-carpet  for ion collection. 
The ions produced by the ESI are transported by the rf-carpet to a quadrupole mass separator (QMS), which was used to select the mass range of interest. 
The selected ions are then guided to a gas-filled rf ion trap where they are accumulated, cooled and ejected as an ion bunch confined to a small phase space. 
Once ejected from the ion trap, the ions enter the MRTOF-MS, where electrostatic mirrors reflect them back and forth, extending their flight path  indefinitely. 
The mirror potentials provide for energy and angle isochronicity. 
As they travel, ion bunches separate as a function of their masses only. 

\subsection{The electrospray ion source}
\begin{figure}[!t]
\centering
\includegraphics[width=\columnwidth]{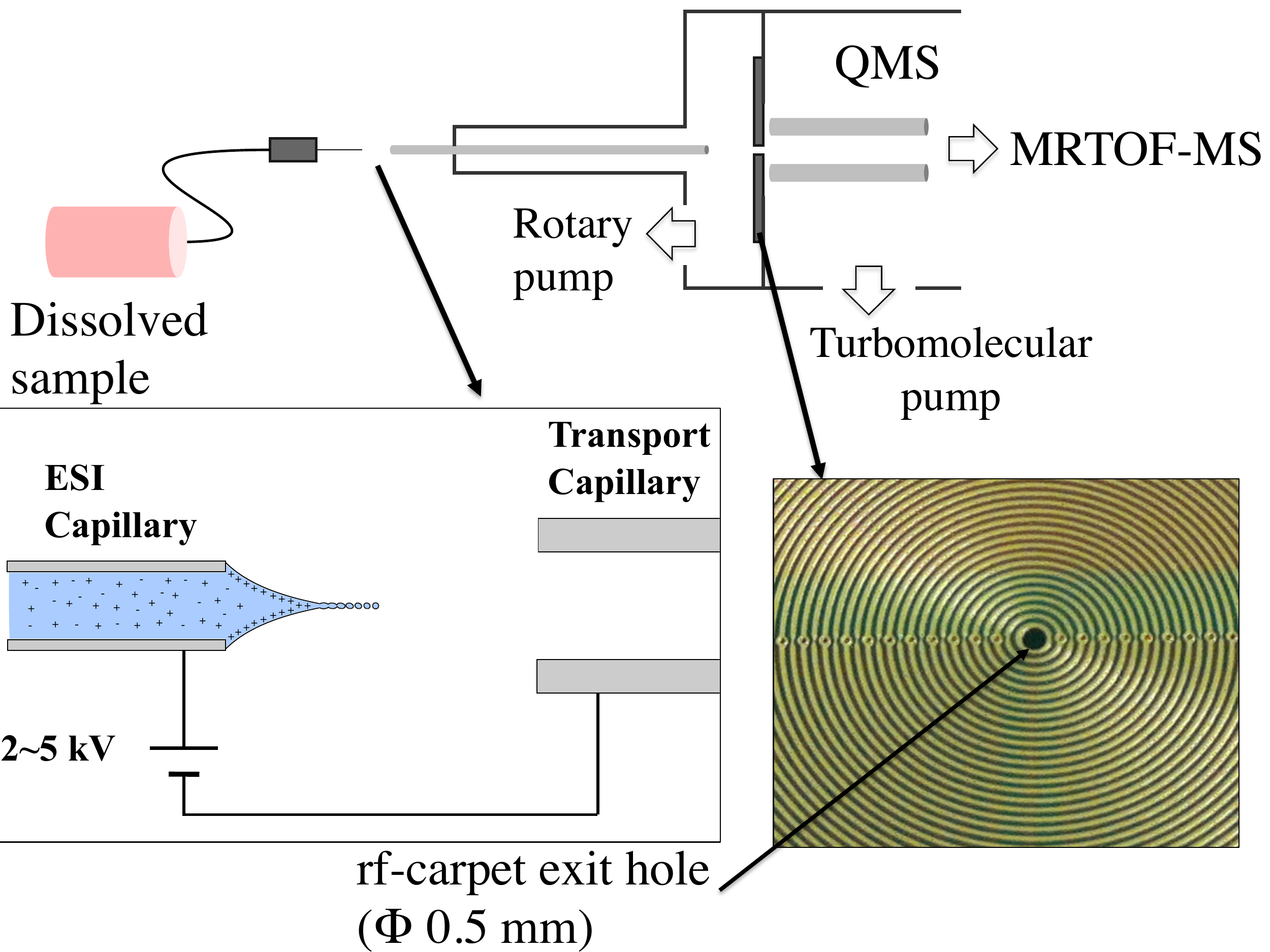}
\caption{(Color online) Schematic overview of the new electrospray ion source, which consists of an electrospray ionization (ESI) system for ion production and an rf-carpet  for ion collection. }%
\label{fig:ESIRF}%
\end{figure}

A schematic overview of the electrospray ion source is shown in  \autoref{fig:ESIRF}. 
To produce molecular ions with electrospray ionization, first a solution is prepared with a specific sample in a solvent with  a trace of formic acid being added to favor positively charged ions. 
In this work, we used three different solutions as listed in  \autoref{tab:Solution}. 
The solution is pushed into an ESI capillary by a motor driven syringe pump. 
A voltage of $2-5\, \rm{kV}$ is applied between the ESI capillary and the transport capillary, which are  separated from each other by $10-30\, \rm{mm}$. 
Due to this short distance, the electric field is as high as  $10^6\, \rm{V/m}$. 
The molecules are polarized near the exposed surface, which forces the liquid to point downfield and form  a so-called Taylor cone \cite{Taylor1965,FernandezdelaMora1994}.  
Near the tip of the Taylor cone the repulsion between positive charges at the surface increases, causing instabilities which form droplets  with approximately the same size as the cone's tip. 
The charged droplets evaporate while drifting toward the transport capillary; the droplets shrink while preserving the charge. 
At a certain radius, known as the Rayleigh limit \cite{Rayleigh1882}, the strong repulsion between charged ions overcomes the surface tension and  Coulomb fission of the droplet occurs.   
Coulomb explosions leads to a jet of ever smaller charged droplets, which undergo cycles of evaporation and Coulomb fission until the droplets are small enough that ion evaporation occurs, forming gas-phase ions \cite{Iribarne1976}.  
Gas flow forces transport the ions into and the through the transport capillary. 
\begin{table}[!htb]
\caption{Solutions used for the production of molecular ions by the ESI. Each solution is prepared by dissolving a sample in solvent (CH$_3$OH or C$_2$H$_5$OH) and formic acid (HCOOH)  mixture. The amount of sample is chosen to provide the desired concentration (Conc.) in the solvent.  Concentrations of formic acid are on the percentage level of the total solution.}
\begin{center}
\begin{tabular}{@{}llll@{}}
\toprule
Sample & Conc. &Solvent   &  HCOOH \\ 
&  (ppm)& & \\
\hline 
C$_6$H$_8$O$_7$ & 200  & CH$_3$OH & 0.1\%\\ 
C$_{17}$H$_{21}$NO$\cdot$HCL& 100  &  CH$_3$OH & 0.1\%\\ 
No sample & & C$_2$H$_5$OH & 10\%\\
\bottomrule
\end{tabular}
\end{center}
\label{tab:Solution}
\end{table}

In conventional electrospray ionization systems, a skimmer is placed behind the transport capillary where a potential difference is applied to pull ions to the high vacuum region. 
In our system, we replace the skimmer with an rf-carpet. 
Just as for the skimmer,  an electric field transports ions to the rf-carpet, where they hoover above  the surface due to an inhomogeneous  rf  electric field, then they are driven by a dc electric field towards the rf-carpet exit hole ($0.5 \, \rm{mm}$ diameter).   

The rf-carpet consists of a planar printed circuit board with 54 ring electrodes, each $0.15 \, \rm{mm}$ with a pitch of $0.3 \, \rm{mm}$ as shown in \autoref{fig:ESIRF}.  
An rf amplitude of up to $300 \, \rm{V}$ has been applied between neighboring electrodes with a frequency up to $4\, \rm{MHz}$ , while a dc gradient of less than $3 \, \rm{V/mm}$ has been superposed between the outmost ring and the center.  
The gas pressure in the rf-carpet chamber is  a few millibar, while in the second chamber the pressure is three orders of magnitude lower. 
The residual gas is mainly air since the ESI system is placed in ambient air.  
When ions reach the rf-carpet exit hole, they are pulled into the second chamber, where a quadrupole mass separator (QMS)  provides an initial mass selection.

\subsection{The accumulation-cooler rf ion trap and the MRTOF-MS} 

To achieve a high mass resolving power, and therefore a precise mass measurement, the ion cloud should have a well-defined energy with a very short-duration time structure. 
To achieve these conditions, a compact gas-filled rf ion trap of novel flat geometry is used to accumulate and cool ions in a few  milliseconds \cite{Schury2009}. 
It consists of a pair of two plates with a $4 \, \rm{mm}$ gap. 
Each plate consists of three segmented strips. 
An rf potential between adjacent strips provides radial confinement of ions, while properly applied dc biasing of the segments creates axial confinement. 
The ion bunch is then transferred from the rf ion trap to the MRTOF-MS wherein it is reflected between the two electrostatic mirrors for sufficient time to achieve a high mass resolving power, $R_m =m/\Delta m = \frac{1}{2}t/\Delta t$. 
A digital PID (proportional-integral-derivative) regulation system is used to stabilize the mirrors' voltages. 
After the flight in the MRTOF-MS, ions are transferred to a micro-channel plate (MCP)  detector placed behind the MRTOF-MS.

\section{Results and discussion} 

First, we describe ion extraction from the new ion source using the rf-carpet. 
The extracted ions were analyzed with the MRTOF-MS to verify the production of molecular isobars.  
 
\subsection{The rf-carpet characteristics}

An electric field transports the ions from the edge of the transport capillary to the rf-carpet. 
The ions are repelled from the rf-carpet surface by an inhomogeneous rf electric field and a dc electric field drives them toward the rf-carpet exit hole. 
According to the well-known Mathieu's equation, the ion motion is stable within a limited parameter space of dc and rf voltages. 
For the rf-carpet, the gas pressure must also be taken into account to find the stability region, which is determined by the two dimension-free parameters $(a - p^2)$ and  $q$ defined as  

\begin{equation}
 a - p^2 = \left(\frac{2 e U_{\text{dc}}}{mr_0^2\omega^2}\right) - \left(\frac{e}{m \mu \omega}\right)^2 
\label{eq:par}
\end{equation}
\begin{equation}
 q = \frac{4 e V_{\text{rf}}}{mr_0^2\omega^2}, %\nonumber
 \end{equation}
where $e$, $m$ and $\mu$  are the electric charge, mass and mobility of an ion, respectively, and $U_{\rm{dc}}$, $V_{\rm{rf}}$ and  $\omega$ are the dc-, rf-amplitudes and the angular frequency applied between adjacent electrodes,  respectively. 
$r_0$ is related to the distance between adjacent electrodes. 
The inhomogeneous rf field can be described by electric pseudo-potential wells with a maximum average effective force approximated in high pressure by \cite{Wada2003}: 

\begin{equation}
\bar{F}_{\rm{hp}}^{\rm{max}} = - \frac{1}{4} m \mu^2 \frac{V_{\rm{rf}}^2}{r_0^3}.
\label{eq:F}
\end{equation}

The rf-carpet pseudo-potential makes a corrugated shape, with the troughs being centered on the electrodes.  
In order for the pseudo-potential approximation to be valid, an ion must spend several oscillations of the field in the vicinity of a single trough.
Efficiently transporting ions to the rf-carpet exit hole by having the required pseudo-potential depth makes the pressure, dc and rf fields critical parameters for stable ion motion.

\begin{figure}[!]
\centering
\includegraphics[width=\columnwidth]{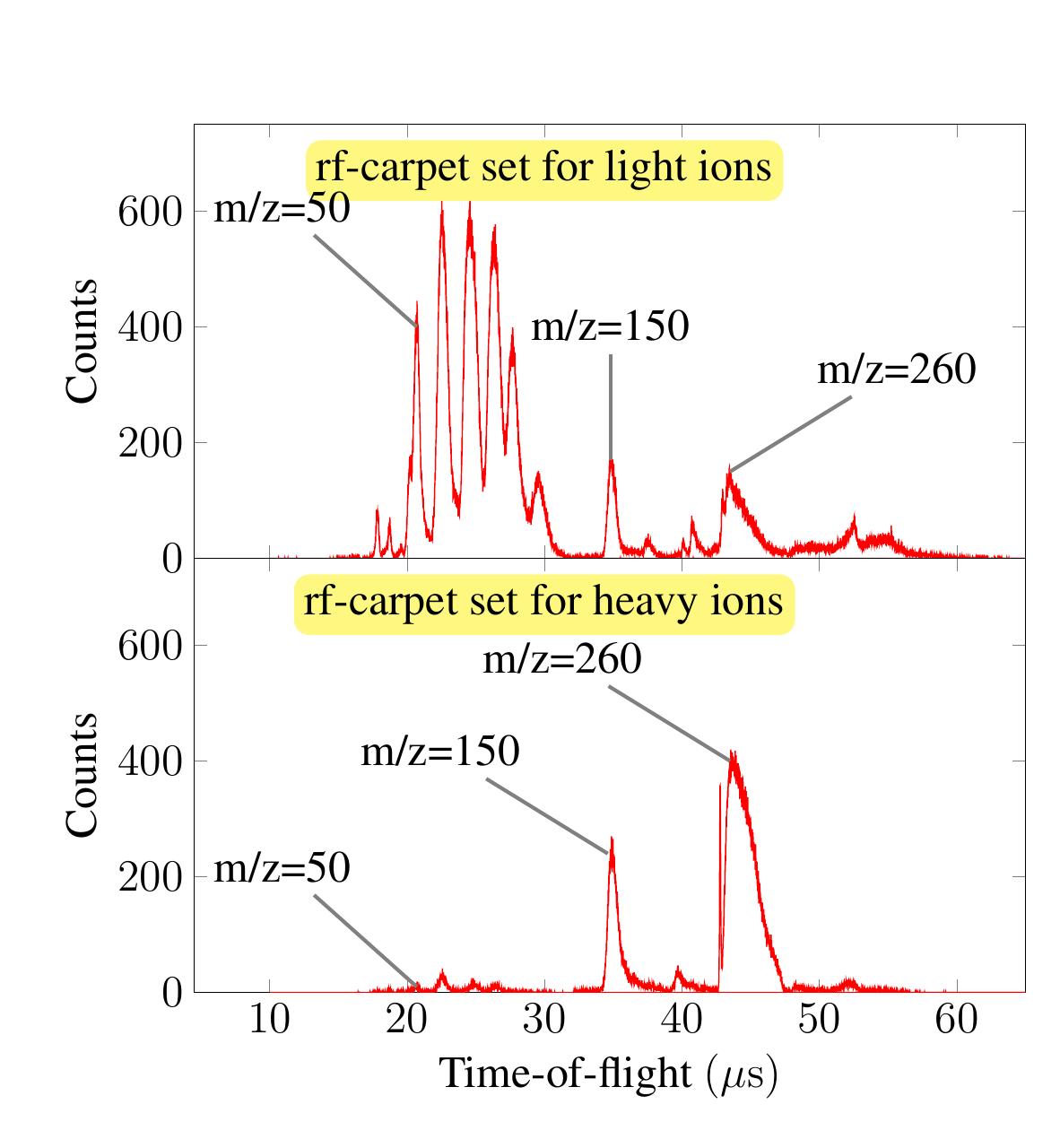}
\caption{Time-of-flight from the trap to the MCP detector behind the MRTOF-MS without any reflection.\textit{(top)} rf-carpet tuned for light ions transmission ($4  \, \rm{MHz}$,  $V_{\rm rf} = 220 \rm{V}$, $V_{\rm dc} = 60 \rm{V}$).\textit{(bottom)}  rf-carpet tuned for heavy ions transmission ($4  \, \rm{MHz}$, $V_{\rm rf} = 60 \rm{V}$, $V_{\rm dc} = 20 \rm{V}$).
}%
\label{fig3}%
\end{figure}

To illustrate the characteristics of ideal transport for a particular mass, we performed tests at  $5  \, \rm{mbar}$ of gas pressure for low- and high-frequency rf fields, varying the rf and dc voltages applied to the rf-carpet ring electrodes. 
The dc voltage, $V_{\rm dc}$, is the voltage applied between the outer and the central  electrodes, where $U_{\rm dc}$ in \autoref{eq:par} can be deduced from $V_{\rm dc}$ \cite{Wada2003}. 
The ions were detected behind the MRTOF-MS after a single pass (without reflections). 
The  time-of-flight (ToF) spectra obtained are shown in \autoref{fig3}. 
The rf-carpet was tuned to ideal conditions for the transmission of light and heavy ions. 
By tuning the rf-carpet parameters (frequency, $V_{\rm rf}$ and $V_{\rm dc}$), a mass range can be chosen as seen in \autoref{fig3}, where low mass ions ($\rm{m/z} < 150$) are drastically reduced when the rf-carpet is set for the transmission of heavy ions.   
\begin{figure}[!]
\centering
\includegraphics[width=\columnwidth]{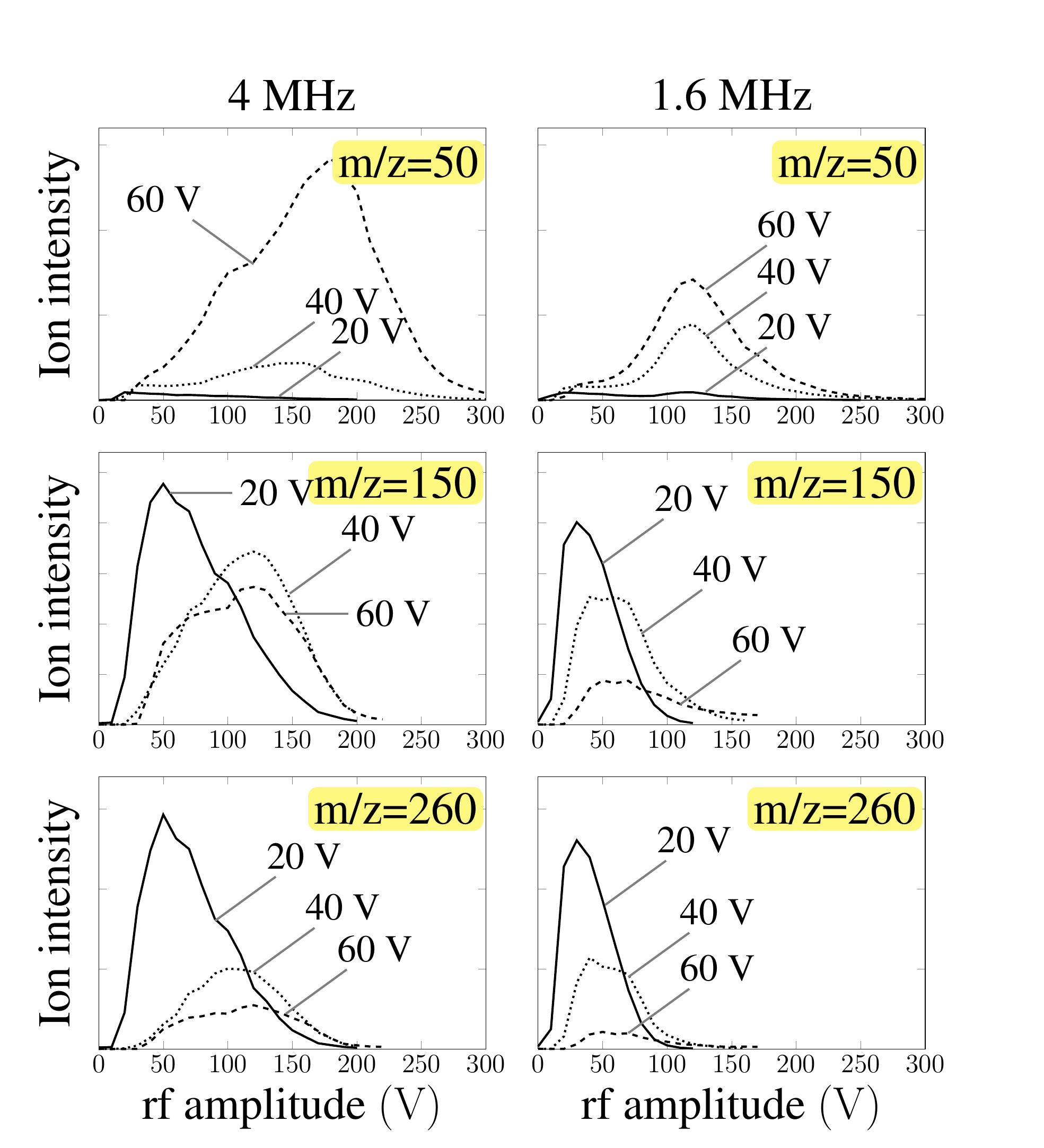}
\caption{Ion intensities for various masses and various dc voltages as functions of the rf amplitude applied to the rf-carpet.}%
\label{fig4}%
\end{figure}
Figure \ref{fig4}  shows the ion intensity for particular examples; $\rm{m/z} = 50$, 150 and 260 from \autoref{fig3}, as a function of rf and dc voltages for frequencies of $4 \, \rm{MHz}$ and $1.6  \, \rm{MHz}$. 
 The ion intensity of low mass ions ($\rm{m/z}  = 50$) decreases at lower frequency, while for high mass ions low-frequency is still sufficient to guide ions to the rf-carpet exit hole. 
Higher rf frequency requires greater rf amplitude. 
Low mass ions require much higher rf amplitude than high mass ions at the same frequency.

Typically, molecular species with low masses have higher mobility than molecular species with higher masses, thus for low frequency they may move between electrodes in less than a few rf periods making the pseudo-potential approximation invalid. 
Therefore, at low frequency, lighter ions may have unstable motion and could be lost before reaching the rf-carpet exit hole, while the motion of heavier ions is slower where the  pseudo-potential approximation is still valid, allowing stable motion. 
At the higher rf amplitudes required for light ions, the pseudo potential well becomes deep, therefore higher dc voltage is required for lighter ions to prevent them from being trapped. 
However, the dc voltage should not be so large as to cause unstable ion motion.

To efficiently transport low mass ions with the rf-carpet, higher frequency and higher dc and rf voltages are required. 
Higher mass ions can be transported using considerably lower frequency, dc and rf amplitudes. 
 
\subsection{The rf-carpet performance} 

The major advantage of using the rf-carpet is efficient collection of ions in buffer gas.  
It can transport ions through small exit hole to high vacuum region which also allows to use simple pumping. 
By using a transport capillary diameter of $0.5 \, \rm{mm}$ a pressure of $1.8 \times 10^{-3}\, \rm{mbar}$ could be reached in the high vacuum region, while in the rf-carpet region the pressure was $4.1\,\,\rm{mbar}$. 
Using a smaller capillary with $0.25\, \rm{mm}$ diameter allowed to achieve a pressure of $1.5\times10^{-4}\, \rm{mbar}$ in the high vacuum region, while in the rf-carpet region the pressure was $0.7\,\,\rm{mbar}$.
These pressures could be reached by using one rotary pump with a pumping speed  of $200 \,\, \rm{l/min} $ in the rf-carpet chamber and one turbomolecular pump with a pumping speed of $300 \,\, \rm{ l/s}$ in the QMS chamber. 

\begin{figure}[!t]
\centering
\includegraphics[width=\columnwidth]{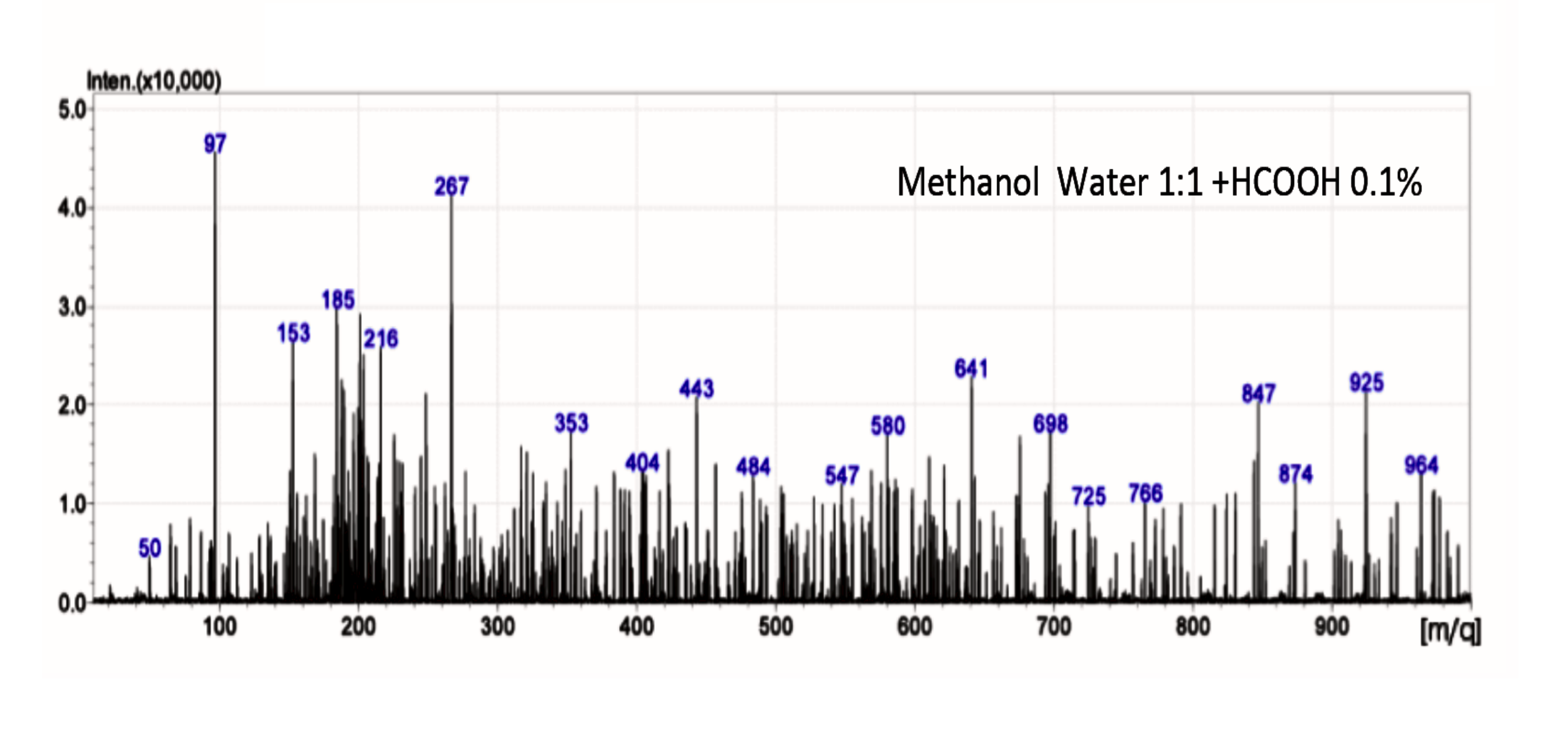}
\caption{Molecular ions produced from a solution of methanol and tap water with a ratio of 1:1 with formic acid ($0.1\%$ of the total volume).   
The spectrum was measured using the Shimadzu 2020MS placed behind the electrospray ionization system. }%
\label{fig:ESISpec}%
\end{figure}

In order to test the efficiency of the rf-carpet, we compare the current impinging on the carpet with the current extracted from the carpet.  
The impinging current was determined by utilizing the carpet itself as a Faraday cup.  
The current extracted from the rf-carpet was determined by utilizing the QMS as a Faraday cup.
For this purpose, a solution of methanol and tap water with a ratio of 1:1 with formic acid ($0.1\%$ of the total volume) was used. 
Part of the large variety of molecular ions produced from this solution is shown in \autoref{fig:ESISpec}, which was measured using the Shimadzu 2020MS placed behind the electrospray ionization system. 
The observed ion current behind the rf-carpet was $\approx15 \%$ of the total ion current observed before the rf-carpet, with the rf-carpet  tuned for low mass region. 
Considering the rf-carpet's limited mass bandwidth and the wide-band production of the ESI, we believe the true transmission efficiency is near unity. 

\subsection{Analysis of heavy molecular ions with the MRTOF-MS}   

\begin{figure}[!tb]
\centering
\includegraphics[width=\columnwidth]{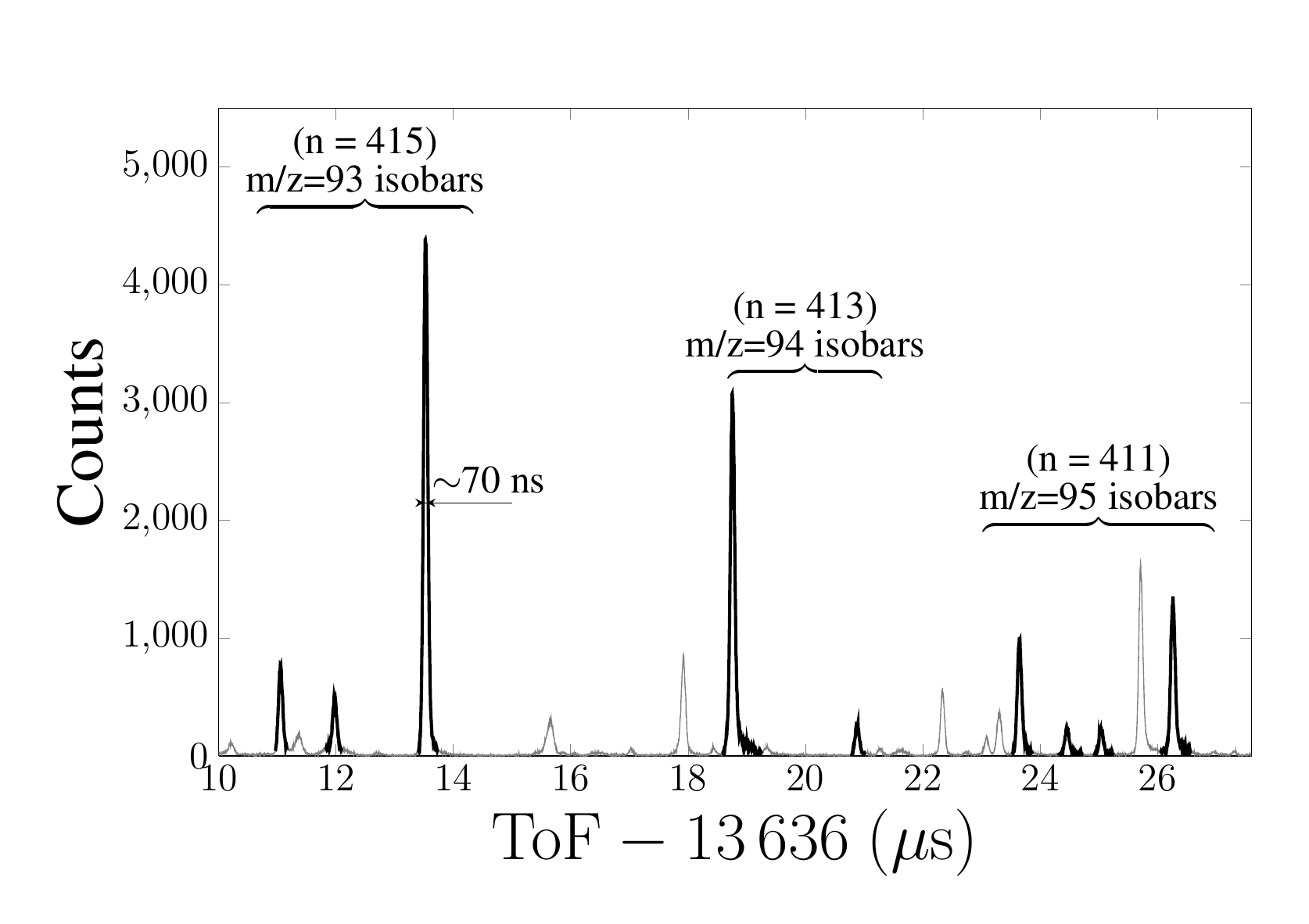} 
\caption{ToF spectrum after a storage time of $13\,629\, \mu s$. The spectrum was obtained from the ethanol solution (\autoref{tab:Solution}). The $\rm{m/z} = 93$, 94 and 95 isobars are well separated. Non highlighted peaks in the ToF spectrum are lighter or heavier molecular ions that come at much higher or lower numbers of revolutions in the MRTOF-MS.} 
\label{fig6}%
\end{figure}

Once the ideal rf-carpet conditions are found, ions are stored in the rf ion trap and transferred to the MRTOF-MS to fly for a time sufficient to separate molecular ions by mass. 
The time between consecutive reflections  in a given mirror (one revolution) can be determined for a known reference ion and scaled for a given ion of interest. 
For this purpose, $^{85}$Rb$^+$ ions from the alkali ion source placed on the other side of the rf ion trap were used as a  marker for low mass ions. 
For higher mass ions, C$_{17}$H$_{21}$NOH$^+$  ions from the solution prepared with diphenhydramine hydrochloride (C$_{17}$H$_{21}$NO$\cdot$HCL  (\autoref{tab:Solution})) were used as a marker.  
After a certain number of reflections, the peak width is minimum -- the time focus condition. 
Thus, the total time-of-flight can be written as: 

\begin{equation}
t = t_0 + n T, 
\label{eq:tofi}
\end{equation}
where $n$ is the number of revolutions of the ions of interest and $t_0$ is to the first order the single pass ToF. 
In a given electrostatic field, light ions fly faster than heavier ions. 
Ions with slightly different masses stored for the same time ($n T$) will have different ToFs. 
Ions with sufficient mass difference may come at different number of revolutions, with heavy ions making lower number of revolutions than lighter ions.  
A typical ToF spectrum at the time focus condition is shown in \autoref{fig6}, where $\rm{m/z}  = 93$, 94 and 95 isobars are well separated.

\begin{figure}[!tb]
\centering
\includegraphics[width=\columnwidth]{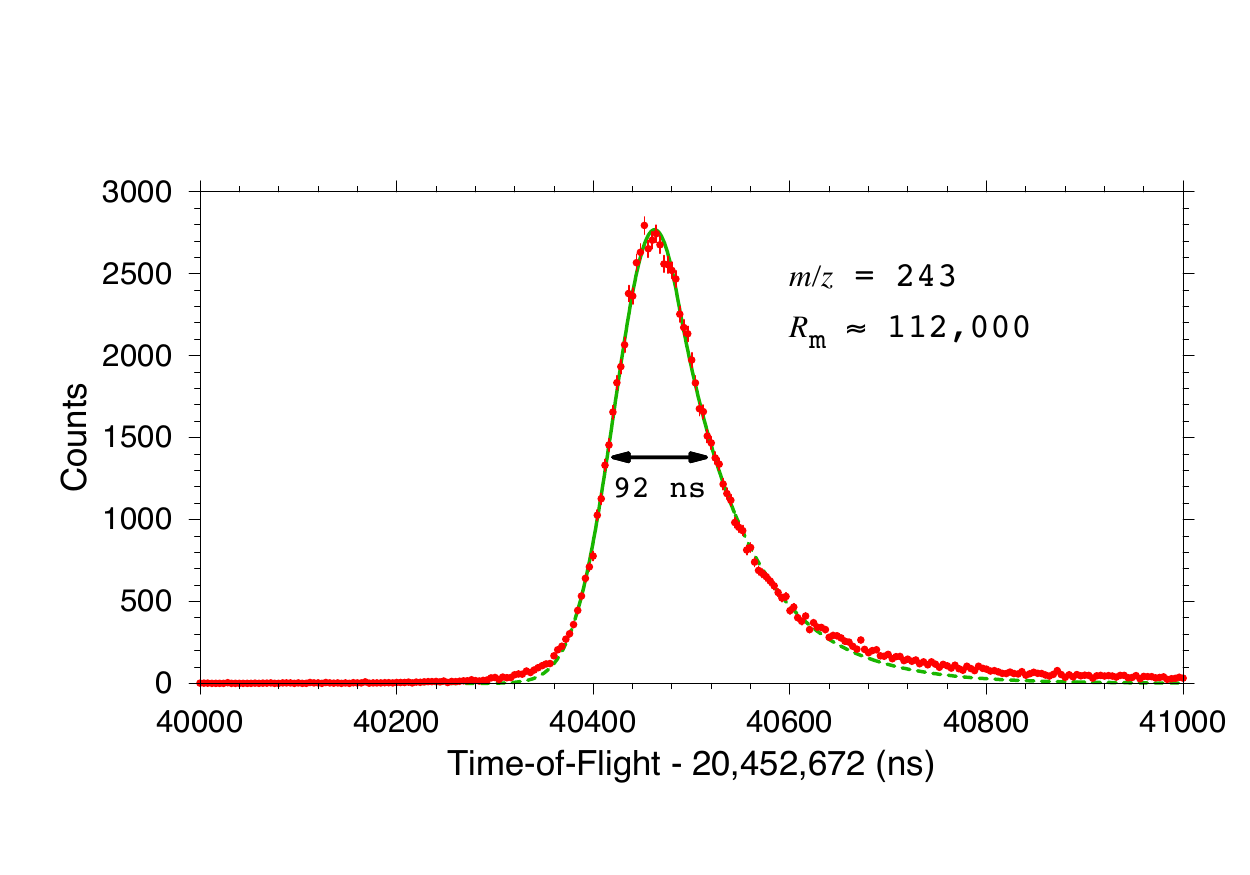} 
\caption{ToF peak of $\rm{m/z} = 243$ fitted with a Gaussian with an exponential tail.} 
\label{fig7}%
\end{figure}

The ToF peaks were fitted with a Gaussian function to obtain the mean ToFs and their uncertainties. 
Figure \ref{fig7}  shows an example of a ToF peak of $\rm{m/z} = 243$ with the fitting function. 
Further details concerning the fitting function of the ToF peaks will be discussed elsewhere.  
Some examples of the results are given in \autoref{tab:Rm}, where molecules with $\rm{m/z} = 243, 244$ and 245 could be used as references for the mass measurement of trans-uranium elements $^{243-245}$Np, $^{243}$Cf, $^{243-245}$Es, $^{243-245}$Fm and $^{245}$Md, which have half-lives of a few milliseconds to minutes. 
A mass resolving power of  $ R_m \approx 100,000$ could be achieved in $13 - 20  \, \rm{ms}$. 
This resolving power could be preserved for heavier masses. 
Due to this high resolving power, the molecular ions from the new ESI ion source could be unambiguously identified even for less abundant molecules as it is the case of $^{13}$CC$_4$H$_6$N$_2^+$  ($\sim 5.4 \%$ natural abundance).

 \begin{table*}[!]
\caption{Time-of-flight, peak width and the mass resolving power obtained for various molecules from the molecular ion source. The peak width $\Delta t$ is the FWHM of the Gaussian fit. }
\begin{center}
\begin{tabular}{@{}lllll@{}}
\toprule
m/z   & Molecules & Time-of-flight (ns) & $\Delta t$ (ns) &  $R_m$  \\ 
\hline \\
 95 & C$_5$H$_5$NO$^+$	& 13,626,528.8(0.3) 	& 70	& 	97,000 \\
                 & C$_4$H$_5$N$_3^+$ & 13,627,327.0(0.6)	& 62 &	 110,000 \\
                 & $^{13}$CC$_4$H$_6$N$_2^+$	& 13,627,916.1(0.6)	& 71 & 	 97,000 \\
                 \vspace{0.3cm}
                 & C$_6$H$_9$N$^+$	& 13,629,136.0(0.3) 	& 67 & 	 102,000 \\
                 
                   \vspace{0.3cm}
243 & C$_{12}$H$_{9}$N$_3$O$_3^+$  &     20,493,142.0(0.5)	& 92	& 	112,000 \\   
         
                    \vspace{0.3cm}
 244 & C$_{12}$H$_{10}$N$_3$O$_3^+$ &   20,482,317.1(1.8)	 & 120 &	 	86,000 \\        

 245 & C$_{10}$H$_5$N$_4$O$_4^+$ & 20,469,137.7(1.4)	& 103 &		99,000 \\
                   & C$_{12}$H$_{11}$N$_3$O$_3^+$ & 20,471,084.7(3.1)	& 138 & 		74,000 \\   
                 
\bottomrule
\end{tabular}
\end{center}
\label{tab:Rm}
\end{table*}

\section{Conclusion}

 We have developed an electrospray ion source coupled to a  radiofrequency carpet for more efficient extraction and simpler pumping. 
The large mass range of the molecular ions produced by this ion source allowed us to successfully test both the rf-carpet and the MRTOF-MS system, which will be used for direct mass measurement of superheavy elements. 
Molecular ions with various masses could be efficiently extracted from the ion source with the rf-carpet.  
With the high mass resolving power of the newly developed MRTOF-MS ($R_m\approx100,000$), the molecular isobars could be easily identified even  with low yield.
The mass resolving power is preserved over a wide mass range. \\

\textbf{Acknowledgments} \\

We would like to thank Drs. S. Davila, G. Eicemann and T. Nishiyama from NMSU, Las Cruces, New Mexico, for their assistance with the ESI techniques. 
This work was supported by the Japan Society for the Promotion of Science KAKENHI (Grant Numbers 2200823, 24224008). 
S. Naimi expresses gratitude to the JSPS Postdoctoral Fellowship for Foreign Researchers. \\

%\bibliographystyle{elsarticle-harv}
%\bibliography{abbreviatednames, actualdatabase} 
\textbf{References}
\bibliographystyle{model1a-num-names}

\bibliography{ESI}

\end{document}